\documentstyle[preprint,aps]{revtex}

\def\B{\beta}

\def \bi{\bibitem}

\def\d{{\rm d}}
 \def\(({\left(}
 \def\)){\right)}
\def\bi{\bibitem}

\def \ov{\over}

\def \b{\beta}
\def\D{\Delta}

\def \d{{\rm d}}
\def \e{{\rm e}}
\def \del{\delta}
\def \nn{\nonumber}
\def \beqna{\begin{eqnarray}}
\def \eeqna{\end{eqnarray}}
\def \beq{\begin{equation}}
\def \eeq{\end{equation}}

\def \ov{\over}

\def \ol{\overline}

\def \b{\beta}

\def \ab2{\alpha\beta^2}

 \newcommand \be {\begin{equation}}
\newcommand \bea {\begin{eqnarray} \nonumber }
\newcommand \ee {\end{equation}}
\newcommand \eea {\end{eqnarray}}

\newcommand \de {\delta}
\newcommand \De {\Delta}

\newcommand \G {\Gamma}
\newcommand \la {\lambda}

\newcommand \lan {\langle}
\newcommand \ran {\rangle}

\begin{document}
\title{Glassy Mean-Field Dynamics of the Backgammon model}

\author{Silvio Franz(*) and Felix Ritort(**)}
\address{(*)  NORDITA and CONNECT\\
     Blegdamsvej 17,\\
     DK-2100 Copenhagen \O
     (Denmark)\\
e-mail: {\it franz@nordita.dk }\\
(**) Departamento de Matematicas,\\
Universidad Carlos III, Butarque 15\\
Legan\'es 28911, Madrid (Spain)\\
e-mail: {\it ritort@dulcinea.uc3m.es}\\}

\date{August 1995}
\maketitle

\begin{abstract}

\end{abstract}
In this paper we present an exact  study of the
relaxation dynamics
of the backgammon model. This is a
 model of a gas of particles in a discrete space
 which presents glassy phenomena as a
result of {\it entropy barriers} in configuration space.

The model is simple enough to allow for a complete analytical treatment
of the dynamics in infinite dimensions.
We first derive a closed equation describing the evolution
of the occupation number probabilities,
 then we generalize
the analysis  to the  study the autocorrelation function.

We also consider possible variants of the model which
allow to study the effect of energy barriers.


\vfill

\narrowtext
\section{Introduction}

The nature of the glass transition is still poorly understood
 \cite{mode,glass}. Under
slow cooling real glasses reach a metastable phase of free energy
larger than that of the crystal phase.
Glasses show a strong slowing down of the
dynamics when the temperature is lowered and the transport coefficients
increase by several orders of magnitude in a  narrow range of
temperatures. It is natural to think that the appareance of high
free-energy barriers is the mechanism responsible for the glass
transition. But free-energy barriers are composed of energy barriers
and entropy barriers.
The question about the relevance of both kind of barriers
in real glasses is of
the outmost importance. Activated jumps of energy-barriers
are
 strongly dependent on temperature.
The typical time
$\tau$ to overcome an energy barrier $\De E$ is
$\tau\sim\exp(\frac{\De E}{T})$ where $T$ is the temperature. This
typical time diverges when the temperature $T$ goes to zero.
Conversely, relaxation times related to  entropy barriers do not depend
directly on the
 temperature.

The simplest way to visualize entropy barriers is the following,
Consider a dynamics in which at each time step the system
can reach a new state with uniform probability;
the typical time to decrease the energy of one unity is $\tau\sim
\frac{\Omega_i}{\Omega_f}=\exp(\D S)$ where $\D S$
is the height of the entropy barrier ($\Omega_i$ stands for the initial
available volume of phase space and $\Omega_f$ stands for the final
volume of phase space with lower energy).

If the effect of energy and entropy barriers is combined one
expects that entropy barriers should affect the temperature activated
relaxation time in its prefactor $\tau\sim
\frac{\Omega_i}{\Omega_f}\exp(\frac{\De E}{T})$.
According to that, the relaxation
time can diverge independently
of the temperature if the phase space volume of lower energy
configurations in the system shrinks to zero during the dynamical
evolution. The idea that
an entropy crisis could be relevant to the glassy transition is very
old \cite{kauzmann,adam}, and it has had  interesting
developements in recent times \cite{kirtirwo,par} in the framework
of mean-field theory of disordered systems. However in the models
studied in \cite{kirtirwo,par}
it is very difficult to disentangle
entropic effects from energetic ones.
To this aim a simple model (the backgammon -BG- model) was recently proposed
by one of us \cite{I} (hereafter referred as I),   in
which energy barriers are completely absent (a diffusive model
with entropy barriers has also been considered in \cite{BM}).
While the model has no thermodynamic transition, it shows a slow dynamics
with strong hysteresis effects and  Arrhenius behavior of the
relaxation time.  The off-equilibrium dynamics of this model was
studied subsequently by us
\cite{II}
(hereafter referred
as II) using an adiabatic approximation,  obtaining fair
good results concerning the relaxation of the energy.
The same approximation has been recently rederived, and slightly refined,
 in a
paper by Bouchaud, Godreche and M\'ezard \cite{BMG}.

In this paper we derive the exact mean-field equation
for the order parameter
for the dynamics of the BG
model, which turns out to be the energy itself.
The techniques
we use are similar to these of \cite{BMG}, however,
the equations we get were not discussed there.
We find that the energy verifies a causal
functional equation with memory. This is at variance with the approximate
treatments where the evolution is described by a Markovian
equation.

In its original formulation, the model does not have any energy
barriers. However, in real systems energy
barriers are present.
The BG model is flexible enough to allow for the introduction (and the
tuning) of energy barriers. This is done by simple modifications
of the Hamiltonian of the system. The formalism developed for the
original model applies in these cases.

In the second section we define the Hamiltonian of the BG model and
the Monte Carlo dynamics we have used to study it.  In the third
section we present some exact results for the behavior of the one-time
quantities (for instance, the energy) and for the two-time quantities
like the correlation function.  The fourth section is devoted to the
study of the effect of energy barriers in the BG model. Finally we
present the conclusions and a discussion of the results.

\section{The BG model and the dynamics}

Let us take $N$ distinguishable particles which can occupy $M$ different
states and let us denote by $\rho=\frac{N}{M}$ the density, i.e. the
number of particles per state. The BG Hamiltonian is defined by,

\be
H=-\sum_{r=1}^M\,\de_{n_r,0}
\label{eq1}
\ee

where $n_r$ is the occupation level of the state $r=1,...,M$, i.e. the
number of particles which occupy that state. The numbers $n_r$ satisfy
the global constraint,

\be
\sum_{r=1}^M\,n_r= N~~~~.
\label{eq2}
\ee

Eq.(\ref{eq1}) shows that energy is simply given by the number of empty
states (with negative sign). We define the occupation probabilities,

\be
P_k=\frac{1}{M}\,\sum_{r=1}^M\,<\de_{n_r,k}>
\label{eq3}
\ee

which is the probability of finding one state occupied by $k$ particles.
The statics of this model in the canonical
ensemble can be easily solved (see (I) and (II)). In
particular one gets the result,

\be
P_k=\rho \frac{z^{k-1} \exp(\beta \de_{k,0})}{k!\exp(z)}
\label{eq4}
\ee

where $z$ is the fugacity and $\beta$ is the inverse of the temperature
$T$ and they are related by the condition,
\be
\rho(e^{\beta}-1)=(z-\rho)e^z~~~~.
\label{eq5}
\ee
expressing that the density is fixed to $\rho$.

The probabilities $P_k$ satisfy the relation $\sum_{k=0}^{\infty}P_k=1$
and they yield all the static observables, in particular the energy
$U=-P_0$.
Several dynamical rules, thermalizing to the Boltzmann distribution,
can be attached to the model. The simplest choice (I) is the
Metropolis single particle dynamics, in which
at each sweep a particle is chosen at random
and it is proposed a move to a new state.
The move is accepted with probability one
if the energy does not increase and with probability
$\exp(-\beta)$ otherwise.

In the mean-field version of the model, the possible
arrival states of the
particles are chosen at random with uniform probability in all the space.
This random motion of the particles allows a complete analytical treatment
of the problem \footnote{The interesting case of a sequential dynamics is more
complicated.}.
Finite dimensional models, where at each sweep
 the particles are only allowed
to move to  neighbours on  a lattice are currently
under study \cite{rito_folla}.

The model has no energy barriers. Consequently there is no
frustration (in the usual sense) and no metastable states. However it was
shown in (I) that the dynamics  is highly non
trivial and a dramatic slowing
down occurs at low temperatures.
This can be qualitatively understood as follows.
Suppose the system is at zero temperature and the dynamics starts
from a random initial configuration of high energy. As the system evolves
 towards the equilibrium
more and more states are progressively emptied and the energy decreases.
Because the average number of particles per occupied state
increases with time (the total number of particles is conserved)
the time needed to empty one more state
 also increases.
The result is that the energy
goes extremely slowly to its equilibrium value.

The dynamical quantities we are interested in are the time-dependent
occupation number probabilities

\be
P_k(t)=\frac{1}{M}\,\sum_{r=1}^M\,\langle\de_{n_r(t),k}\rangle
\label{eq6}
\ee
($E(t)=-P_0(t)$)
and  the two time energy-energy correlation function
\cite{I},

\bea
C_E(t,s)=& &\frac{\frac{1}{M}\sum_r\delta_{n_r(t),0}\delta_{n_r(s),0}
-E(t)E(s)}{-E(s)(1+E(s))}
\nn\\
\equiv & &
{P(n_r(t)=0,n_r(s)=0)-P_0(t)P_0(s)\ov P_0(s)[1-P_0(s)]}
\,\,\,\,t\ge s
\label{eq7}
\eea
At finite temperature, when $t,s>>t_{eq}\sim \exp(\b)/\b^2$ (see
reference (II) and also section III.B) this function is time
translationally invariant. In the regime in
which both times
$t,s$ are much less than $t_{eq}$, and at all times
at zero temperature,  the system is off-equilibrium,
time-translation invariance does not hold, and the correlation function
displays aging (see \cite{I}).

\section{Mean-field equations for the dynamics of the BG model}

In this section we derive exact mean-field
equations for the Monte Carlo
dynamics of the BG model.  First we
adress the dynamical problem associated to the one-time probability
distributions $P_k(t)$. These probabilities generate an
infinite  hierarchy of Markovian equations which can be closed
in terms of  the only quantity $P_0(t)$.
 Then  we will study the
two-time correlation functions  in a
similar way.  For simplicity, we will
restrict all the future computations to the case $\rho=1$ (i.e $M=N$),
 the generalization to an arbitrary density being very simple.

\subsection{Dynamical equations for $P_k(t)$}

The  purpose of
 this section  is to write the dynamical evolution equations
for the probabilities $P_k(t)$ and, in particular, for the internal
energy $E(t)=-P_0(t)$. An elementary
Monte Carlo move consists in a random selection of one particle (hence, the
probability to select a particular departure state $d$
 is $n_d/N$ where
$n_d$ is the occupation level of that state) and moving it to a randomly
selected arrival state $a$ with uniform
probability independent of the occupation level $n_a$. One Monte carlo
step (our unity of time) consists of $N$ elementary moves.
In the elementary move there are several processes
which contribute to the variation of $P_k(t)$.
In appendix A we write explicitly the balance equations,
the result  is:

\be
 {\d P_k(t)\ov \d t}=
(k+1)(P_{k+1}-P_{k})+P_{k-1}+
P_0(e^{-\beta}-1)(\de_{k,1} - \de_{k,0} - kP_k + (k+1)P_{k+1})
\label{eq10}
\ee
where the time index for the probabilities $P_k$ has been omitted.
The previous equation holds for $k\ge 0$ with $P_{-1}=0$.
In particular for $k=0$ we obtain the equation studied in (II),

\be
\frac{\partial P_0}{\partial t}=P_1(1-P_0) - e^{-\beta} P_0 (1-P_1)
\label{eq11}
\ee
The hierarchy was closed in (II) assuming fast relaxation on the surfaces
of constant energy, and slow variation of the energy itself. In this
condition eq.(\ref{eq11}) was solved assuming
$P_k(t)={\exp[{\b(t)\del_{k,0}-z(t)}]}{z(t)^{k-1}\ov k!}$ with $\b(t)$ and
$z(t)$ related at all times by eq.(\ref{eq5}).

Here we study the hyerarchy (\ref{eq11}) with the method of the generating
function, that we define as

\be
G(x,t)=\sum_{k=0}^{\infty}\,x^k\,P_k(t)
\label{eq12}
\ee

A similar approach was also used in \cite{BMG} where the adiabatic
approximation of (II) was rederived and improved\footnote{The
technique of the generating function in the study of the dynamics
has also been applied
to some mean-field spin glass models \cite{BPPR}.}.

{}From the equation (\ref{eq10}) it is easy to check that the $G(x,t)$
satisfies the partial differential equation,

\be
\frac{\partial G(x,t)}{\partial t}=(x-1)[G(x,t) + \la(t)
- (1+\la(t))\frac{\partial
G(x,t)}{\partial x}]
\label{eq14b}
\ee

with $\la(t)=P_0(t) (e^{-\beta}-1)$.
Eq. (\ref{eq14b}) is a  non linear partial differential
equation, the nonlinearity
is contained
in the dependence of $\lambda$ on $P_0(t)=G(0,t)$.

The equilibrium solution $G_{eq}(x)$ is easily obtained
from equations (\ref{eq4}) and (\ref{eq12}),

\be
G_{eq}(x)=\frac{e^{\beta}-1+e^{zx}}{z\,e^{z}}
\label{eq15b}
\ee
 and one can check that this is consistent with eq.(\ref{eq14b}).

The previous partial differential equation can be implicitly
solved to get $G(x,t)$ as a functional of $\lambda$.
The details are presented in the Appendix B, we give here the result

\be
G(x,t)=\e^{(x-1)D(t,0)}G_0(1+(x-1)B(t,0))+
(x-1)\int_0^t \d s \lambda(s) B(t,s) \ \e^{(x-1)D(t,s)}
\label{aa}\ee
where we have written
\beqna
B(t,s)&=&\exp\left({-\int_s^t \d v \ [1+\lambda(v)]}\right)
\nn\\
D(t,s)&=&{\int_s^t \d v \ B(t,v)}
\eeqna

and $G_0(x)=G(x,0)$ is the initial condition at time $t=0$.
Setting $x=0$ in (\ref{aa}) we get a closed equation for
$P_0(t)$  which reads

\be
P_0(t)=
\e^{-D(t,0)}G_0(1-B(t,0))
+(1-e^{-\B})\int_0^t \d s P_0(s) B(t,s) \ \e^{D(t,s)}
\label{bb}
\ee

The previous equation, although
strongly non-markovian is {\it causal}, as the l.h.s. depends on
the values of $P_0(s)$ for $s\le t$. It has  a unique solution
that can be found numerically  with good
precision, discretizing the time and integrating
it step by step.
The evaluation of the
previous expressions gives the full solution of the BG model as far as the
one-time dynamical quantities are concerned.

The solution of (\ref{aa}) is explicit at infinite temperature ($\b=0$).
In this case $\la(t)=0$ and the solution of $G(x,t)$ simplifies,

\be
G(x,t)=e^{(1-e^{-t})(x-1)}\,G_0((x-1)e^{-t}+1)
\label{cc}
\ee
It is not surprising that
at infinite temperature the system goes exponentially fast to the
equilibrium (with
relaxation time equal to 1).
At infinite temperature the equilibrium probabilities eq.(\ref{eq4}) are
given by $P_k=\frac{1}{k!e}$, the energy being $E=-P_0=-\frac{1}{e}$. If
we start from the initial condition in which all particles occupy the same
state ($P_0=1,\,P_k=0,k>0$) then we have $G_0(x)=1$. From eq.(\ref{cc})
we obtain the time evolution of the energy,

\be
E(t)=-G(0,t)=-e^{e^{-t}-1}
\label{dd}
\ee

We studied numerically  the solution of (\ref{bb}) at $T=0$. In
figure 1 we display the result for the energy, starting from
the initial condition $P_k(0)=1/(e k!)$ at time 0 (i.e. $G_0(x)=e^{x-1}$).
For comparison
we plot the results of the Monte Carlo simulations and of the
adiabatic hypothesis of (II) with the same initial
condition.\footnote{The adiabatic hypothesis  gives better results if
the integration is started at latter times.}

\subsection{The Correlation Function $C_E(t,s)$}

In this section we investigate the behavior of the
energy-energy correlation functions (\ref{eq7}).
We proceed  in a similar
way as we have done for the occupation probabilities.
We need to study the joint occupation
probability in a given site $r$ at two different times
$t,s$ ($t>s$),  $P(n_r(t)=0,n_r(s)=0)=P(n_r(t)=0|n_r(s)=0)P_0(s)$.
The correlation function  eq. (\ref{eq7}) can be written as
\be
C_E(t,s)={P(n_r(t)=0|n_r(s)= 0)-P_0(t)\ov 1- P_0(s)}.
\ee
We now write a set of equations that allow to study
$P(n_r(t)=0|n_r(s)= 0)$.

Let us define the probabilities
\be
\nu_k(t,s)=P(n_r(t)=k|n_r(s)=0)
\ee
i.e. the occupation number probabilities in the set
$S_s$ of states which are empty at time $s$.
In general, it is possible
to restrict the  balance equations that led to (\ref{eq10})
to any subset $S$ of the whole space.
Irrespectively of $S$ the result is:

\beqna
\frac{\partial \nu_k}{\partial t}=
& &\nu_{k-1}-\nu_{k}+[(k+1)\nu_{k+1}-k\nu_k][1-P_0(1-e^{-\beta})]
\nn\\
& &-(\de_{k,1}-\del_{k,0})[\nu_0(1-P_1)+\nu_1 P_0](1-e^{-\beta}).
\label{nuk}
\eeqna
In particular if the set $S$ is the whole space $\nu_k=P_k$ and
we get back to (\ref{eq10}).

Of course the initial
conditions depend on the set under study.
For the set $S_s$ we are interested to, we must choose

\be
\nu_k(s,s)=\del_{k,0}.
\ee

In terms of the generating function
\be
\G(x,t,s)=\sum_{k=0}^\infty x^k \nu_k(t,s)
\ee
eq.(\ref{nuk}) reads

\be
\frac{\partial \G}{ \partial t}=
(x-1)[\G-(1-P_0(1-e^{-\beta}))\frac{\partial \G}{ \partial x}
-(\nu_0(1-P_1)+\nu_1 P_0)(1-e^{-\beta})]
\label{gamma}\ee
with  condition at time $s$
\be
\G_s(x)\equiv \G(x,s,s)=1.
\ee
Note that if we suppose to know the $P_k(t)$ then the
eq.(\ref{nuk},\ref{gamma})
are linear. Obviously, if one considers the
set $\ol{S}_s$ complementary to $S_s$, and its respective generating
function $\ol{\Gamma}$, it holds the equality:
$P_0(s) \G(x,t,s)+(1-P_0(s))\ol{\G}(x,t,s)=G(x,t,s)$.

Defining
\beqna
& &\mu(t,s)=[\nu_0(t,s)(1-P_1(t))+\nu_1(t,s)P_0(t)](1-\e^{-\b})
\nn\\
& &B(t,s)=\exp\left({-\int_s^t \d v \ (1-P_0(t)(1-\e^{-\b})) }\right)
\nn\\
& &D(t,s)=\int_s^t \d v \ B(t,v)
\eeqna
we find
\be
\G(x,t,s)=
\e^{(x-1)D(t,s)}-
(x-1)\int_s^t \d u \mu(u,s) B(t,u) \ \e^{(x-1)D(t,u)}
\label{bbb}\ee
which depends implicitly on $\nu_0$ and $\nu_1$.
In order to find a closed
system we have to consider eq.(\ref{bbb}) and its $x$-derivative in
$x=0$
\beqna
& &\nu_0(t,s)=1+ \int_s^t \d u \ \left[
-\nu_0(u,s)[1-(1-P_1(u))(1-\e^{-\b})]+\nu_1(u,s)\right]
\nn\\
& &
\nu_1(t,s)= \int_s^t \d u \
\left[
\mu(u,s) B(t,u)\e^{-D(t,u)}(D(t,u)-1)
\right]
+\e^{-D(t,s)}D(t,s)
\label{volt}
\eeqna

The system (\ref{volt}), if one assumes known  the probabilities
$P_k(t)$, consists in a vectorial
linear Volterra equation  of second kind for $\nu_0$ and $\nu_1$
which can in all generality be
integrated  numerically,
 and in some particular case also analytically.

The simplest case is the  equilibrium
at finite temperature. In that case, $P_k\equiv P_k^{eq}$ and
the various functions  appearing in (\ref{volt}) are time traslation
invariant. In these conditions eq.(\ref{volt}) can be
solved in Laplace transform. Simple algebra, and the
formula (see e.g. \cite{tricomi})
\be
\int_0^\infty d t \ exp(-a \exp(t)-E)=a^{-E} \ \gamma(p,a)
\ee
($\gamma$ is the incomplete gamma function),
 shows that $\nu_0(E)$,  the Laplace transform of $\nu_0(t-s)$, is given
by
\be
\nu_0(E)={A(E)+{z-1\ov z}(1-E A(E))\ov 1-\left[{(z-1)e^z\ov (z-1)e^z +1}
+{z-1\ov z}E\right](1-E A(E))}
\label{lap}
\ee
where we have expressed all the equilibrium quantities in terms
of the fugacity $z$ (see eq. (\ref{eq5})),
$E$ is the Laplace variable conjugated to time, and
\be
A(E)={1\ov e^z z^{zE-1}}\int_0^z u^{zE-1}e^u.
\ee
$\nu_0(E)$, as it should, has a pole in $E=0$ with residue $P_0$ coresponding
to $\nu_0(t)\to P_0$ for large time. Poles on the real negative $E$ axis
correspond to exponential relaxation modes. The largest relaxation time
is given by minus the inverse of the value of $E$ in the pole closest
to the origin. This can be obtained explicitely for large $\b$, where
$z\approx \b-\log(\b)$ is large, from the asympotic expantion of
$E A(E)$ for small $E$
\be
E A(E)\approx
e^{-z}+E.
\ee
The result is simply $E_{pole}\approx -e^{-z}$ and correspondingly
$\tau_{max}\approx e^z\approx exp(\beta)/ \beta$.

In the off-equilibrium regime the integration of (\ref{volt}) can be
performed
numerically.
In fig. 2 we show the result of the integration for $T=0$ for different
values of $s$ ( i.e. different waiting times) compared to the
Monte Carlo results.
Although we did not try very sophisticated algorithms, with standard
ones \cite{num_rec}, we were able to reach enough large times to
see the scaling
behavior $C_E(t,s)=f((t-s)/s)$ observed numerically in I.
It would be interesting to see if equation (\ref{bbb}) could be solved
with the aid of some simple approximation as it is the case
for the energy (II and \cite{BMG}).

\section{The effect of energy barriers}

The BG model has no energy barriers and hence there is no finite
temperature thermodynamic phase transition. In real glasses energy
barriers are always present and it can be instructive to understand
their effect when combined with entropy barriers. One can easily
modify the Hamiltonian (\ref{eq1}) to include
energy barriers. In this paper we have focused
on two different ways. In the  first, we have
considered interaction
 between the different states, introducing an
energy gain when groups of states are
simultaneously empty. This interaction term is enough to make
appear a finite temperature thermodynamic transition, but
metastability and frustration are absent and the system monotonically
reaches the ground state at zero temperature. In the other,
we introduced metastable
configurations in the dynamics. In this case the system fails to reach
the ground state at zero temperature while thermodinamically
there is no finite-temperature phase transition.

\subsection{The $p$-states model}

The simplest way we can introduce interaction among
different states in the model is the following, consider the quantity
\be
M[\{ n_r\}] ={1\ov N}\sum_{r=1}^N (\del_{n_r,0}-1/e).
\ee
Any Hamiltonian of the form
\be
H=N F(M[\{ n_r\}])
\label{hf}
\ee
with $F$ gentle enough,  is a good candidate for a mean-field model.
We did not try a systematic study of the form (\ref{hf}) for generic
$F$, but we concentrated to the class of monomials, where
\be
H_p=-\frac{1}{N^{p-1}}\left(\sum_{r=1}^N(\de_{n_r,0}-{1}/{e})\right)^p
\label{23}
\ee

For $p=1$ this model reduces to the BG model.
For larger values of $p$ there is
interaction between different states. The ground
state of this model is the same as the one of
 the BG model (all particles
occupying the same state) and there are no energy barriers at zero
temperature.
A careful study of the thermodynamics of this model shows that for any
$p>1$ there is a first order phase transition from a
completely disordered phase with $M=0$ for $T>T_c$ to an `ordered' phase
with $M\ne 0$ for $T<T_c$. This leads to the curious situation that the
completely disordered state is dynamically stable at all temperatures
but at $T=0$. This can be understood by a simple argument.
Suppose to start the dynamics in a random initial condition
and consider a sweep that lead to the filling of an empty
state. The energy variation in this process is
\be
\del H=-{1\ov N^{p-1}} \left[ \left(\sum_r(\del_{n_r,0}-{1\ov e})-1 \right)^p
-\left(\sum_r(\del_{n_r,0}-{1\ov e}) \right)^p \right]\simeq
{1\ov N^{p-1}} p \left(\sum_r(\del_{n_r,0}-{1\ov e}) \right)^{p-1}.
\ee
But according to our hypothesis $\sum_r(\del_{n_r,0}-{1\ov e})$
is a random variable of order
$\sqrt{N}$, correspondingly $\del H\sim N^{-(p-1)/2}$ and the acceptance
rate
\be
\e^{-\b \del H}\sim \e^{-\b /N^{(p-1)/2}}
\ee
For finite temperature and large $N$ all the moves are accepted and the
energy in average never decreases. In other words
the
statistic of configurations is not changed by the dynamics.
A crossover is found for $\b\sim N^{(p-1)/2}$ showing that the relevant
scale of temperature for the dynamics is different
from that of the statics. Right at zero temperature, where
only the sign of the energy  change and not the magnitude
 matters,  the dynamics
of the model   coincides for any $p$  with the one of the conventional
case $p=1$.

\subsection{The effect of metastability}

The $p$-states model has no metastability at zero temperature. We
want to study here  a simple model where metastability is present but without
interaction. In the BG model the ground state is reached by emptying
progressively more and more states. To empty a given
 state at a certain time
$t$ it is necessary
to pass in a configuration where a unique particle occupies that state.
We then  consider the following model,

\be
H=\sum_{r=1}^N\,(-\de_{n_r,0}\,+\,g\de_{n_r,1})
\label{30}
\ee

where $g$ is positive constant and we have the usual constraint
eq.(\ref{eq2}). At zero temperature
the transition $n_r=2 \ \to \ n_r=1$ is forbidden, hence
energy barriers are present in the model.
More general models
are obtained  including in the Hamiltonian all possible terms of the type
$\de_{n_r,k}$

\be
H=-\sum_{r=1}^N\,\sum_{k=0}^{\infty}g_k\de_{n_r,k}=-\sum_{r=1}^N\,g_{n_r}
\label{31}
\ee

We focus here to the case (\ref{30}).
The statics of this model is easily solved. We obtain the free energy,

\be
\beta f=\log(z)-\log(e^z+e^{\beta}-1+z(e^{-\beta g}-1))
\label{32}
\ee

and the fugacity is related to the temperature $\frac{1}{\beta}$ by the
$g$-independent relation eq.(\ref{eq5}). The equilibrium probabilities
$P_k$ (see eq.(\ref{eq3})) are given by,

\be
P_k=\frac{z^{k-1} \exp(\beta\de_{k,0}-\beta g \de_{k,1})}{k!
(e^z+e^{-\beta g}-1)}
\label{33}
\ee

The dynamics of this model is expected to be substantially different to
that of the BG model at least at very low temperatures. Concretely, at
zero temperature the ground state is the same as for the BG case but
there is a large number of metastable states (for instance, half states
empty and half of the states with two particles). It is easy to show
that for each value of $E$ between $E=-1/2$ and the ground state $E=-1$
there exists a metastable configuration with that energy. Then we expect
the value of the energy extrapolated to infinite time to depend strongly
on the initial configuration. In order to minimize the energy we have to
maximize $P_0$ and minimize $P_1$. While the maximization of $P_0$ is a
process where entropy barriers are dominant (this is the reason why the
BG model defined as $E=-P_0$ is interesting) this is not the case for
minimizing $P_1$ where entropy barriers are absent. Then, independently
of the initial configuration we expect that $P_1$ will go to zero in the
 exponentially fast for large times. In these conditions, we do not
 expect
that the adiabatic solution of (II) can give a good approximation
of the dynamics. This
 approximation was based on the
fact that in the BG the surfaces of constant energy are connected,
a situation which does not hold here.

However the dynamics of this model
can be directly solved as in the BG case.
Skipping all the details we find for the generating function:

\bea
\frac{\partial G(x,t)}{\partial t}=(x-1)\Bigl(-(1+\la(t))\frac{\partial
G(x,t)}{\partial x} + (1+2 P_2(e^{-\beta g}-1))G(x,t)+\\
\la(t)-2 P_2 (e^{-\beta g}-1)(e^x-P_0 (x-1)(1-e^{-\beta
(1+g)})\Bigr )
\label{36}
\eea

where $\la(t)=P_0(t) (e^{-\beta(1+g)}-1)$. Observe
that eq.(\ref{36})
depends only on the probabilities
$P_0(t)$ and $P_2(t)$.  The solution  is
 more complicated to that of  eq.(\ref{aa}), however it can be found.
In figure 3 we show the result of the numerical
integration of (\ref{36}) for the energy at $T=0$ compared with the
 Monte Carlo simulations, starting from a completely random initial
condition. The energy seems to converge to a value $\lim_{t\to\infty}E(t)=
-.564$, a result that it would be interesting to  derive  analytically.
It is under current study the  finite temperature
dynamics, where we expect the effect of the energy and entropy
barriers to combine to give rise to a dynamics slower than that
of the BG model.

\section{Conclusions}

In this paper we have derived  the exact mean-field equations
of the dynamics of the Backgammon model. This has been achieved
through the study of the single site occupation number probability
for which a hierarchical set of equations hold. With the method of the
characteristic function, we have derived a closed functional
equation for the energy. This, although non-Markovian,
 has a causal character and can be integrated
step by step discretizing the time. The non-Markovian character of the
evolution equation, suggests that history dependent effects should
be observable in the system. However the analysis of II, where
the evolution of the energy was described by an approximate equation,
shows that even in subtle phenomena as hysteresis cycles
in cooling-heating processes, history dependent effects are very small.
This should reflect in the fact that the memory kernels that
appear in the equation for the energy are  short range in time.

The method of the generating function also allowed us to derive also
a system of linear Volterra equations describing
the evolution of the energy autocorrelation function.
The numerical solution of these equation confirmed the aging
behaviour found in I. It would be interesting to derive analytically
the scaling
$C_E(t,t')=f(t'/t)$.

In the last section we have derived the mean-field theory
for a model where entropic and energetic barriers are combined.
We have seen that at temperature zero, starting from a random configuration
the system fails in finding the ground state. For future work
it is left the study of this model for finite temperature.

Non linear equations with memory appear in
phenomenological glass theory
under the name of Mode Coupling Theory \cite{mode}.
Mode coupling equations appear naturally in the
mean-field treatment of the
dynamics of disordered \cite{cuku,frame}
or quasi-disorderd systems \cite{FH}, in off-equilibrium
situation they involve a set of coupled
integral equation for the two time auto-correlation function
and its conjugated response function. The most striking
manifestation of the importance of memory effects in off-equilibrium
mode coupling theory is in the
aging behaviour of the response function \cite{cuku,frame}.

 Structural glasses are generally
classified as strong glasses (Arrhenius behavior of the relaxation time)
or fragile glasses (Vogel-Tamman-Fulcher behavior of the relaxation time).
In this classification the BG model is a strong glass. Polymer glasses are
fragile glasses which show strong aging effects in its physical properties
\cite{glass}.
It would be desiderable to know from experiments if there is a correlation
between the fragility of glasses and its aging properties. This could
shed light on the role of energy barriers in the mechanism of the
glass transition. We believe  that only entropy barriers cannot
yield aging effects in the response function. In this framework a more
detailed
study of the BG model with metastability (as presented in the last section)
at finite temperature could be instructive. In particular, the study
of the relaxation time as a function of the temperature
and the existence of aging due to the presence
of energy barriers.

\begin{center}
{\bf Acknowledgements}
\end{center}
We would like to thank J.P. Bouchaud, E.Follana and M. M\'ezard for
interesting discussions. (F.R.) acknowledges Ministerio de Educacion
y Ciencia of Spain for financial support.

\section{Appendix A}
In this appendix we derive the evolution equation for
the probability $P_k(t)$.
Define as $N_k(t)$ the number of states occupied by
$k$ particles ($P_k(t)=N_k(t)/N$).

 The processes leading to a variation of $N_k$ can
be classified in this way:
\begin{itemize}
\item{{\it Process A$+$}: arrival of a particle in a state
with $k-1$ particles.}
\item{{\it Process A$-$}: departure of a particle from a state
with $k$ particles.}
\item{{\it Process B$+$}: departure of a particle from a state
with $k+1$ particles.}
\item{{\it Process B$-$}: arrival of a particle in a state
with $k$ particles.}
\end{itemize}
Note that in a sweep the processes above are not mutually
exclusive, so, for example the contemporary
occurency of $A+$ and $B-$ lead to no variation in $N_k$.
At each Monte Carlo sweep three independent random variables are extracted:
a departure state $d$ with probability $n_d/N$, an arrival state $a$
with probability $1/N$ and an acceptance variable
\be
x=
\left\{
\begin{array}{lll}
1 & {\rm with \;\;\; prob.} & \e^{-\b}
\\
0 &{\rm with \;\;\; prob.}&1- \e^{-\b}
\end{array}
\right.
\ee
 In terms of these
variables the variation in $N_k$ in each process is given by:
\begin{itemize}
\item{{\it Process A$+$}: $\ \ \de_{n_a,k-1}
[1-\de_{k,1}(1-\de_{n_d,1})\del_{x,0}]$}
\item{{\it Process A$-$}: $\ \ -\de_{n_d,k}[1-\de_{n_a,0}\del_{x,0}
+ \de_{k,1}+\de_{k,1}\de_{n_a,0}\del_{x,0}]$}
\item{{\it Process B$+$}: $\ \ \de_{n_d,k+1}[1-\de_{n_a,0}\del_{x,0}
+\de_{k,0}\de_{n_a,0}\del_{x,0}]$}
\item{{\it Process B$-$}: $\ \ -\de_{n_a,k}
[1-\de_{k,0}(1-\del_{n_d,1})\del_{x,0}]   $}
\end{itemize}
The contribution in the different processes can be easily understood,
for example in process A$+$ we must have $k-1$ particles in the
arrival state. If $k=1$ and $n_d>1$ the move imply an energy cost, and
is accepted only if $x=1$.

Summing  all the contribution
and averaging over $p$,  $a$ and $x$  we find:
\bea
& &
\lan
N_k(t+\del t)-N_k(t)\ran =
N[P_k(t+\del t)-P_k(t)]\equiv {\d P_k(t)\ov \d t}=
\nn\\
& &(k+1)(P_{k+1}-P_{k})+P_{k-1}+
P_0(e^{-\beta}-1)(\de_{k,1} - \de_{k,0} - kP_k + (k+1)P_{k+1})
\label{A10}
\eea

Very similar considerations  lead to (\ref{nuk})
if one restricts the balance equation to a subset
of the whole
space.

\section{Appendix B}

In this Appendix we obtain the solution of eq.(\ref{eq14b}).
We perform the change
of variables $(x,t)\to(u,t)$ where $x-1=e^{u+\int_0^t\d s (1+\lambda(s))}
=e^u B(t,0)$.
 In terms of the new
variables  eq.(\ref{eq14b}) reads,

\be
\frac{\partial \hat{G}}{\partial t}=
e^u B(t,0)(\hat{G}
+\la)
\label{ap2}
\ee

where $\hat{G}(u,t)=G(x(u,t),t)$
This is a linear differential equation which can be readily solved

\be
\hat{G}(u,t)=e^{e^u\int_0^t\, \d s B(s,0)} F(u)
+
 e^u
\int_0^t \d s \ \lambda(s)B(s,0)e^{e^u\int_s^t\, \d v B(v,0)}
\label{ap3}\ee

where $F$ is an arbitrary function.

Going back to $(x,t)$ and imposing
the initial condition we get  eq.(\ref{aa}).

\vfill
\newpage
{\bf Figure Captions}
\begin{itemize}

\item[Fig.~1]
The decay of the energy at zero temperature starting
from a completely random configuration at time $t=0$.
We compare the numerical solution
of (\ref{bb}) (full lines) with the Monte Carlo simulations for $N=10^5$ and
the integration of the adiabatic equation of II with the same initial
condition.

\item[Fig.~2]
The correlation function at zero temperature
as a function of $(t-t_w)/t_w$ for different $t_w$
($t_w=10,30,100,300$).
We take it as a good indication for the $t/t_w$ scaling at large times.

\item[Fig.~3]
The correlation function at zero temperature for $t_w=10$ compared with
the montecarlo data for $N=10^5$.

\item[Fig.~4]
Energy vs. time in the model with energy barriers
(theory + Monte Carlo data at $N=10^5$),
 starting from a random configuration
at time zero. We observe exponential decay to $E=-.564$.

\end{itemize}

\end{document}